# SNRAware: Improved Deep Learning MRI Denoising with SNR Unit Training and G-factor Map Augmentation

**Hui Xue[1], Sarah M. Hooper[2], Iain Pierce[4], Rhodri H. Davies[3,4], John Stairs[1], Joseph Naegele[1], Adrienne E. Campbell-Washburn[2], Charlotte Manisty[4], James C. Moon[4], Thomas A. Treibel[4], Peter Kellman[1,2]\*, Michael S. Hansen[1]\***

1. Microsoft Research, Health Futures, Redmond, WA, USA
2. National Heart, Lung and Blood Institute, National Institutes of Health, Bethesda, MD, USA
3. Institute of Cardiovascular Science, University College London, London, UK
4. Barts Heart Centre, Barts Health NHS Trust, London, UK

\*These authors contributed equally as senior authors to this work.

## Corresponding author:

Hui Xue

Microsoft Research, Health Futures

Building 99, Room 4941
14820 NE 36th St
Redmond
WA 98052

Email: xueh@microsoft.com

# SNRAware: Improved Deep Learning MRI Denoising with SNR Unit Training and G-factor Map Augmentation

**Key points:**

1. We propose a new training method for deep learning MRI denoising. The key innovation is to integrate quantitative noise distribution information from SNR Unit reconstruction and g-factor augmentation to improve model performance.

2. This training method is agnostic to model architecture and validated on 14 different models from two backbone types including both transformer and convolutional layers.

3. We trained models on a large dataset of 96,605 cine series and validated the models extensively on in -and out-of-distribution tests. In-distribution, we show that the proposed method improves performance on a test dataset of 3,000 cine series. Out-of-distribution, we show that models trained with the proposed method on 100% cardiac cine data generalize to different imaging sequences, dynamic contrast variations, and anatomies.

**Summary statement:**

SNRAware is a model-agnostic approach to train MRI denoising models that leverage information from the image reconstruction process, improving performance and enhancing generalization to unseen imaging applications.

**Abbreviations**

deep learning = DL, signal-to-noise ratio = SNR, magnetic resonance imaging = MRI, Magnetization Prepared Rapid Gradient Echo = MPRAGE, balanced steady-state free precession = B-SSFP, two-chamber = CH2, three-chamber = CH3, four-chamber = CH4, short-axis stack = SAX, Fast Low Angle Shot MRI = FLASH, standard deviation = SD, peak signal-noise-ratio = PSNR, structural similarity index measure = SSIM, contrast-to-noise ratio = CNR

**Abstract**

**Purpose**

To develop and evaluate a new deep learning MR denoising method that leverages quantitative noise distribution information from the reconstruction process to improve denoising performance and generalization.

**Methods**

This retrospective study trained 14 different transformer and convolutional models with two backbone architectures on a large dataset of 2,885,236 images from 96,605 cardiac retro-gated cine complex series acquired at 3T. The proposed training scheme, termed SNRAware, leverages knowledge of the MRI reconstruction process to improve denoising performance by (1) simulating large, high quality, and diverse synthetic datasets, and (2) providing quantitative information about the noise distribution to the model. In-distribution testing was performed on a hold-out dataset of 3000 samples with performance measured using PSNR and SSIM, with ablation comparison without the noise augmentation. Out-of-distribution tests were conducted on cardiac real-time cine, first-pass cardiac perfusion, and neuro and spine MRI, all acquired at 1.5T, to test model generalization across imaging sequences, dynamically changing contrast, different anatomies, and field strengths.

**Results**

The in-distribution tests showed that SNRAware training resulted in the best performance for all 14 models tested, better than those trained without the proposed synthetic data generation process or knowledge of the noise distribution. Models trained without any reconstruction knowledge were the most inferior. The improvement was architecture agnostic and shown for both convolution and transformer attention-based models; among them, the transformer models outperformed their convolutional counterparts and training with 3D input tensors improved performance over only using 2D images. The best model found in the in-distribution test generalized well to out-of-distribution samples, delivering 6.5× and 2.9× CNR improvement for real-time cine and perfusion imaging, respectively. Further, a model trained with 100% cardiac cine data generalized well to a T1 MPRAGE neuro 3D scan and T2 TSE spine MRI.

**Conclusions**

An SNRAware training scheme was proposed to leverage information from the MRI reconstruction process in deep learning denoising training, resulting in improved performance and good generalization properties.

# Introduction

Recent work has shown that deep neural networks can restore signal from low signal-to-noise ratio (SNR) magnetic resonance images (MRI) better than earlier, conventional denoising methods (1). For low-SNR imaging applications, such as low-field MRI (2,3), diffusion imaging (4), acquisition with higher parallel imaging acceleration (5), and cardiac dynamic imaging (6), deep learning (DL) denoising models can restore diagnostic image quality and increase clinical value.

Deep learning based denoising for MRI can be achieved with either supervised or self-supervised training. The former requires paired samples of noisy and clean data. However, curating large amounts of paired training data can be difficult, requiring time-consuming acquisition of high-SNR data or repeated measurements and special imaging protocols that lengthen the scanning sessions. This data curation can be very challenging for applications with lower intrinsic SNR, e.g. imaging with high undersampling rates to avoid motion. Self-supervised training (7–10) methods have been proposed to circumvent this limitation by proposing training pipelines that only rely on noisy images. In this category, Noise2Noise (8) learns the noise pattern from a pair of noisy images. Noise2Void (9) requires only one noisy image and learns to predict a blind-spot pixel from its surroundings. Noise2Fast (10) also needs only one noisy image and predicts a small patch from neighboring patches. However, these methods are slow at inference due to the need to run training on every dataset and their performance is inferior compared to supervised training (11). More recent developments in the self-supervised domain utilize diffusion generative models to improve quality. Examples include the denoising diffusion model for diffusion MRI (4) and score-based diffusion sampling (12).

Previously published work did not explore potential performance gain from noise information in the training. Moreover, the training was mostly performed on limited datasets up to a few thousand samples with a specific contrast and imaging sequence (4,8,10,12). The resulting models may not be robustly transferrable to other imaging setups, especially for applications with intrinsically low SNR, where the starting quality is poor and denoising is more challenging. It is, however, exactly in those cases where denoising is mostly needed. In MRI, the noise distribution can be derived from the reconstruction process. In this work, we investigate how to leverage this information to enhance denoising performance and generalizability.

We propose a new training scheme for MRI denoising that we term *SNRAware*. Uncorrelated starting noise after the noise pre-whitening (13) is augmented by the real g-factor maps to create spatially varying noise. Noise correlation augmentation is introduced to mimic operations like *k*-space filters, phase oversampling and image resizing. SNR unit reconstruction (14) is applied through the noise generation process to maintain the unity noise level and help model learning. A g-factor augmentation method is proposed to compute g-factor maps for acceleration factors ranging from R=2 to R=8, even if the data are not collected at those specific accelerations. This method contrasts with prior work, which typically trains models on normalized signal levels, not noise. During training, paired low- and high-SNR images are not required and noise is computed on-the-fly and added to the high SNR data.

We trained denoising networks on a very large dataset. The value of g-factor augmentation, realistic MR noise and training with SNR levels was evaluated by comprehensive ablation comparison in the in-distribution test. We hypothesize that the improvement is agnostic to model architecture and investigate this by comparing 14 different model architectures. Furthermore, we hypothesize this noise-centric training method enables models to generalize to unseen applications, such as different tissue contrasts, imaging sequences, field strengths, and anatomies.

## Materials and Methods

### *Data collection*

In this retrospective study, all training data were retro-gated cardiac cine imaging data acquired on 3T clinical scanners (MAGNETOM Prisma, Siemens AG Healthcare). A balanced steady-state free precession (B-SSFP) sequence was used with typical acquisition parameters listed in Table 1. All retro-gated cine imaging was acquired with R=2 acceleration. Common cardiac views, such as two-chamber (CH2), three-chamber (CH3), four-chamber (CH4), and short-axis stack (SAX) were scanned. The raw *k*-space signals were saved for the following reconstruction. The data used in this study was not utilized in prior publications.

Data was from the NIH Cardiac MRI Raw Data Repository, hosted by the Intramural Research Program of the National Heart, Lung and Blood Institute. All data were curated with the required ethical and/or secondary audit use approvals or guidelines that permitted retrospective analysis of anonymized data without requiring written informed consent for secondary usage for the purpose of technical development, protocol optimization, and quality control. All data was fully anonymized and used in training without exclusion.

Table 1 summarizes the training and test datasets. A total of 96,605 cine series (2,885,236 images) from 7,590 patients were included for training. Typically, every slice series includes 30 phases. The in-distribution test set consisted of 3000 retro-gated cine series. There was no overlap between training and test sets.

To test model generalization for different imaging sequences, contrast, field-strength and anatomy, four out-of-distribution tests were presented. Ten real-time cine slices were scanned at the medial short axis locations, with the B-SSFP contrast but different sequence parameters than training. To evaluate generalization over dynamic contrast changes, ten free-breathing first-pass perfusion scans were acquired. For other anatomies, a neuro T1 MPRAGE 3D scan for a R=2×2 acceleration and a spine T2 TSE multi-slice 2D scan for R=2 was acquired. All generalization datasets were acquired at 1.5T (MAGNETOM Aera, Siemens AG Healthcare).

Further, phantom scans were acquired at 1.5T with R=2 and 4 acceleration and the standard FLASH (Fast Low Angle Shot MRI) readouts.

### Training method

SNRAware leverages the MR reconstruction process to aid model training in two ways: (1) to generate low SNR data, and (2) to provide information on the noise distribution to the network. Each of these contributions is detailed below. Figure 1 illustrates the training data generation process. Figure 2 outlines the training scheme and model design.

#### Training data generation with g-factor augmentation

The training data was acquired with undersampling rate of R=2, which allowed a reconstruction with minimal or no g-factor related noise enhancement. To help generalize to higher acceleration, we implemented a g-factor based data augmentation scheme in which real g-factor maps for higher acceleration R=3 to 8 were calculated and used to apply realistic noise. With parallel imaging, the noise becomes spatially varying, scaled by the g-factor (18,19). This noise amplification is the result of the ill-posed inversion of the calibration matrix and varies from scan to scan. After reconstruction, the noise SD for a pixel location $p$ is $\boldsymbol{g}(p)$. $\boldsymbol{g}$ increases dramatically for higher acceleration (as shown in Figure 2 and 3).

Our process to compute different g-factors is shown in Figure 1a. The auto-calibration or full-sampled $k$-space lines were used to compute 2D GRAPPA coefficients. The $k$-space GRAPPA convolution kernels were converted to the image domain unmixing coefficients (17,20). The g-factor maps were computed as the sum of squares of the unmixing coefficients. Although the acquired scan had an acceleration factor of two, g-factor maps for other undersampling factors were estimated by computing corresponding unmixing coefficients. In training, one of the g-factor maps is chosen at random and used to amplify white, complex noise via pointwise multiplication.

The spatial noise distribution is altered by other reconstruction steps like $k$-space filter or zero-filling resizing in a way that creates spatial noise correlation. A training data augmentation process was created to vary the training data noise correlation in a way that mimics commonly used reconstruction steps. White noise was sampled for each training image with a noise sigma randomly selected from zero (not adding noise) to a prescribed maximal level (32.0 used in this study). The sampled noise was first amplified by a g-factor map and then altered using a $k$-space filter (Gaussian filter, sigma selected from [0.8, 1.0, 1.5, 2.0, 2.25]). Partial Fourier filter was applied with a probability of 0.5 (tapered Hanning filter (21); partial Fourier sampling ratio [1.0, 0.85, 0.7, 0.65, 0.55]). Reduced resolution was mimicked by masking out high frequency samples (ratio: [1.0, 0.85, 0.7, 0.65, 0.55]). All these operations were independently sampled for readout and phase encoding directions.

By random selecting starting noise sigma, acceleration, and $k$-space filters, the augmentation procedure produces a wide range of spatially varying SNR that closely resembles what would be observed in a variety of imaging protocols. Figure 1b shows generated noisy samples with different SNR, illustrating a wide range of SNR was achieved by changing both sampled noise standard deviation (SD) and g-factor map. Supplemental Movie 1 shows the corresponding movies for these samples.

*Providing noise distribution information to the network*

To aid in the denoising task, we provide information about the noise distribution to the network. We use SNR unit (14) reconstruction to reconstruct all training and test data. This reconstruction method carefully scales the noise SD to be unity and maintains this noise scaling through the reconstruction pipeline. We hypothesize that this method would aid the denoising model by reducing the variation in noise distributions which the model must learn. To perform SNR unit reconstruction, pre-scan noise-only data was acquired before every imaging scan (14). The noise readouts were used to compute the covariance matrix and perform noise pre-whitening on the imaging readouts (16,17). The noise SD was scaled to be 1.0 by compensating for the equivalent noise bandwidth for every receiver coil or channel. The imaging data with unity noise went through FFT, parallel imaging GRAPPA reconstruction (18), and coil combination to produce the final complex images. The noise scaling was kept constant throughout all these steps (14). The complex images were finally resized with zero-filling to the target matrix size.

We applied similar techniques to keep the noise SD constant when generating synthetic noise for our training data. Given a high SNR image $\boldsymbol{I}$ which was reconstructed while maintaining unit noise variance scaling through all signal processing steps, except parallel imaging unmixing, and a corresponding native g-factor $\boldsymbol{g}$, the corresponding SNR unit image is $\boldsymbol{S} = \boldsymbol{I}/\boldsymbol{g}$. Generated, correlated noise $\boldsymbol{n}$ with a selected variance $\sigma^2$ and augmented according to the scheme outlined above was added to the SNR unit image to create a noisy sample: $S_n = (S + n * g_{aug})/\sqrt{\sigma^2 + 1}$, where $g_{aug}$ is a g-factor map computed in g-map augmentation outlined above. The ratio $1/\sqrt{\sigma^2 + 1}$ accounts for original unity noise and added noise and returns the image scaling to unit noise variance except for the noise amplification introduced by parallel imaging unmixing. Every training pair thus consists of a clean sample $S$ and noise augmented sample $S_n$. The clean image has unity noise, and the noisy image has spatial varying noise multiplied by $g_{aug}$.

This process is illustrated in Figure 2a. In addition to providing the network images with unity noise, we also provide the g-factor map to the network as an input stacked along the channel dimension. This directly provides the network with information about the spatial amplification of noise in the image.

Sample code and detailed explanations can be found in previous publications and tutorials (13,16,17,20) for the noise pre-whitening, SNR unit scaling, and how to compute pixel-wise g-factor maps from parallel imaging calibration. Gadgetron framework (https://github.com/gadgetron/gadgetron) provides an open-source, high-performance implementation.

***Model and training***

The inputs to all models were 5D tensors [B, C, T/S/D, H, W] for Batch, Channel, Time or Slice or Depth, Height, and Width. This representation provided flexibility to support different imaging formats. For example, for the input cine series, the 3rd dimension was time. For a 3D neuro scan, it was depth or slice. The g-factor map was concatenated to real and imaginary part of image tensors, so C was 3 for complex training; if only the magnitude image was used, C was 2. The model output a tensor with the same shape except the channel dimension was 2 for complex and 1 for magnitude images respectively.

We evaluated 14 architectures. As shown in Figure 2b, these models were based on two adapted backbone types: HRnet (22) and U-net (23). Both backbones use multi-resolution pyramids to balance computational complexity with the ability to recover small image features by maintaining a full resolution path. Each network consists of multiple blocks. Each block contains several cells. Every cell includes normalization, a computing layer, and a mixer. Different models were instantiated by configuring different computing layers. Both transformer layers and convolution layers were tested. The transformer layers include layers inspired by the Swin (24), ViT (25), and more recent CNNT (11) models, where input tensors are split into patches across T/S/D, H, W and attention is computed over patches. For Swin, we split the input image into patches and apply attention over local and shifted windows. For the ViT, attention is global over all patches. The CNNT cells do not patch the image and instead apply attention in the T/S/D dimension. We also tested convolution layers (referred to as "Conv" blocks), which do not patch the image but apply standard convolution. All cells include three layers, except CNNT-large with six layers. The ViT2D and Conv2D models were further trained by employing 2D patching and attention, or 2D convolution, operating over H and W, not across frames. These many configurations enable us to assess SNRAware over transformer models, convolutional models, 2D and 3D models, as well as multiple backbone configurations. More information on model building is provided in Appendix E1 (supplement).

The loss was the sum of Charbonnier loss (26), MR perpendicular loss (27) which was designed to match complex values, VGG-perceptual loss on magnitude (28), and the gradient loss which was computed as the L1 difference of intensity gradient between ground-truth and predicted tensors.

The dataset was split with 95% for training and 5% for validation. A fast second order optimizer, Sophia (29), was used with the one-cycle learning rate scheduler (30) and cosine annealing. The peak learning rate was 1e-5, betas were 0.9 and 0.999, and epsilon was 1e-8. The training lasted 80 epoch, and the final model was selected as the one giving the highest performance on the validation set. All models were implemented using PyTorch (31) and training was performed on a cluster with 128 AMD MI300X GPUs, each with 192GB RAM. Data distributed parallelization was used across multiple GPU cards to speedup training.

### *Evaluation*

#### *In-distribution test*

The same noise generation process was used to generate low-SNR images for the test dataset of 3,000 series. Resulting data was fed into the trained model. The peak SNR (PSNR, computed as $10 \cdot log_{10}(\frac{MAX^2}{MSE})$) and structural similarity index measure (SSIM) (32) were computed on model outputs against the clean ground-truth. $MSE$ is the mean square difference. $MAX$ is the maximal value of image pixels. Since the image signals here are floating values and noise level is unity, $MAX$ is set to 2048.0, as the SNR above this high threshold would be highly unlikely.

PSNR and SSIM are reported for 14 tested models (two backbone types; layer types: CNNT, CNNT-large, Swin3D, ViT3D, ViT2D, Conv3D, Conv2D). Following ablations were further performed. *Without g-factor map:* Models were trained without the g-factor map supplied as the extra input channel and the inference did not take in g-factor map. *Without MR noise*: Training included g-factor map, but generated noise was not transformed by filters and noise was added to the high-SNR images using uncorrelated white noise. *Magnitude training without imaging knowledge*: Simulating reconstruction information not being available and operating on magnitude images such as one would find in DICOM images. Here, the training was performed by excluding g-factor maps, MR realistic noise and using the magnitude images. The channel dimension size was thus 1.

*Generalization tests*

High SNR ground-truth images were not available for the imaging data acquired with higher acceleration. Here, the SNR gain was estimated with the Monte-Carlo simulation method (14) by repeatedly adding a fixed amount of noise to the input data for N=64 times. The noise level in the model outputs was measured by computing the standard deviation across repetitions. The SNR increase was measured by the reduction of noise SD. Regions-of-interests (ROIs) were drawn in myocardium and blood pool. The SNR and the CNR (contrast-to-noise ratio, as $\frac{2\times(Signal_{blood} - Signal_{myo})}{(noise_{blood}+noise_{myo})}$) were measured.

A paired t-test was performed and a *P*-value less than .05 was considered statistically significant.

# Results

Figure 3 shows the phantom test results. Three measurements were made on the g-factor maps for R=2 and R=4 scans. For R=4, the g-factor was elevated in the center of the water phantom corresponding to areas of elevated noise in the image. Results from denoising with two of the proposed models (HRnet-CNNT and Unet-Conv3D) are shown in the figure. Combining g-factor and MR noise both improved the performance. The best SNR was achieved with the proposed scheme where most information from reconstruction was incorporated into the training. The worst performance was seen when training was done without knowledge of reconstruction. The transformer based CNNT model outperformed Conv3D in this test.

Table 2 summarizes the in-distribution test results. The ablation was repeated for all 3D or transformer models. Because the 2D models without ablation were not competitive to 3D training, we did not attempt ablation there. Uniformly, the proposed scheme gave the best PSNR and SSIM. Either removing g-factor or realistic MR noise from training led to worse performance. Training without reconstruction knowledge led to the worst performance. For both backbone types, CNNT-large surpassed other models. HRnet-CNNT-large and Unet-CNNT-large offered the best PSNR and SSIM, with the former achieving highest performance across all models evaluated. Comparing 3D models to their 2D counterparts (e.g., ViT3D vs. ViT2D, Conv3D vs. Conv2D), the 3D models were superior.

The model with highest scores, HRnet-CNNT-large, was used in the generalization tests. Table 3 lists the SNR and CNR results for cardiac generalization tests.

Figure 4 shows an example of real-time cine results for R=5. The real-time cine imaging used an bSSFP readout similar to the training data with, but with higher acceleration and different protocol parameters (such as bandwidth, matrix size, resolution). The proposed training gave the best quality with improved SNR across the field of view. The improvement with reconstruction knowledge was significant (P<1e-5, Table 3). The differences between model outputs and raw images show removed noise resembles a similar pattern to the g-factor map. Amplified noise was removed to a lesser extent when g-factor map was not used in training, which is also visualized in Supplemental Movie 2. ROIs were drawn in blood pool and myocardium for all N=10 cases. The mean SNR improvement of the proposed method was 5.2× for blood pool and 3.5× for myocardium, significantly higher than all ablations. The blood pool and myocardium CNR increased 6.5×. Supplemental Movie 3 presents more real-time cine examples.

Figure 5 gives the results for perfusion imaging, where a contrast bolus was injected and passed through the heart. This represents a larger departure from the training data distribution, given the dynamic changes in image contrast. The model trained with image reconstruction knowledge successfully improved SNR over the contrast passage. Like the previous sample, the noise amplification was removed over the field of view. Models without full imaging reconstruction information were inferior compared to the proposed training. This is also quantitatively verified by the SNR measurements in Table 3. Using the proposed training method, the mean SNR increase was 3.0x for blood pool and 3.7× for myocardium. The CNR increased by 2.9×. Supplemental Movie 4 shows the corresponding movies.

Figure 6 shows results on two other anatomies. Both cases are out-of-distribution, with differences in imaging sequences, resolution, contrast and anatomies. There was no head or spine data in the training dataset; still, the model was able to noticeably increase SNR. The zoomed images show well-preserved details. Figure 6a measures the white and gray matter intensities before and after the model, and we observe the SNR gain did not alter the contrast. Figure 6b is a T2 TSE spine scan with high spatial resolution (0.76mm$^2$). The model improved SNR noticeably for vertebratae, discs, spinal cord and cerebrospinal fluid. Supplemental Movie 5 shows the neuro results in the sagittal view. Spine results are shown in Supplemental Movie 6.

## Discussion

This study proposes a new training scheme, *SNRAware,* which integrates knowledge from the image reconstruction process into deep learning denoising training for MRI. SNR unit reconstruction was employed to produce unit noise level images, which simplifies the task of augmenting the training data with realistic noise distributions. Geometry (g)-factor maps were appended as model input data, thus supplying quantitative information about spatially varying noise amplification. We proposed an augmentation method, which computes real g-factor maps for R=2 to 8, while only requiring the training data to be acquired with a single acceleration factor. MR realistic noise was generated on-the-fly to lower the SNR of ground-truth in a manner that closely resembles the noise distribution in the reconstructed images with higher acceleration factors and varying $k$-space filtering effects. This method removes the need to have paired high and low SNR images for every acceleration factor and filter configuration, which is would not be practical for imaging with every acceleration or resolution.

We tested the training scheme on 14 model architectures from two backbone types, covering both transformer-based layers and convolution. The evaluation was performed on phantoms, in vivo in-distribution data of N=3000 cine series, and four out-of-distribution data. Results show that integrating reconstruction information into the training pipeline consistently improved model performance. At higher acceleration factors, such as the R=5 real-time cine test, the proposed method effectively corrected the g-factor noise amplification, while models trained without g-factor maps did not. Finally, we hypothesized models trained to recognize MR noise distribution may generalize to unseen imaging applications, which was supported by experiments on perfusion imaging with dynamically changing contrast and on neuro and spine scans.

The design of training and model architecture generalizes over different data dimensionalities by processing the tensors in the shape of *[B, C, T/S/D, H, W]*. For the 2D+T case, like the cine series, the 3$^{\text{rd}}$ dimension is time. For a 3D scan like the neuro T1 MPRAGE, it is the second encoding dimension or depth. For the spine scan which acquired 15 2D slices, it is the Slice dimension. We show that the models denoise each of these formats, despite being trained only on 2D+T data. This makes it practical to combine different training data (e.g. 2D+T, 3D and multi-slices) into one model training session, which may further improve the model generalization.

Previous studies have proposed using g-factor maps into MR denoising. In a recent paper (33), 23 neuro T2 scans were collected and used to train a CNN model. The g-factor maps were used in loss computation, instead of as model input as proposed in this study. Another study (34) provided g-factor maps as input with the low SNR images into the training. In that study CNN model was trained with 2000 T2 neuro data and simulated noise. Tests were in-distribution on other neuro images. Our study trained models on larger datasets for both transformers and convolution architecture. The impact of noise amplification caused by g-factor and noise correlation caused by raw filter and other steps were separately tested in our study and the out of distribution validations here are more extensive. This study emphasizes a noise centric view of the denoising training can improve the generalization of trained model to unseen imaging applications.

There are limitations to this study. Firstly, noise pre-whitening is required to perform SNR unit scaling. This requires noise calibration data, acquired by turning off RF pulses and sampling. Such data can be acquired in a separate acquisition or a few leading readouts in an imaging scan, but critically the data must be preserved and used in image reconstruction to perform noise scaling. Some commonly used MRI raw datasets (such fastMRI data (35)) does not come with noise scans and thus cannot be appropriately scaled. Secondly, this study implemented noise augmentation using a concrete set of processing steps (like k-space filter, partial Fourier etc.) to generate realistic MR noise distributions. The framework would need to be extended if new processing steps, which may alter noise distribution, are introduced. Finally, we demonstrated out of distribution generalization only on a limited set of examples. Further evaluation is needed to understand how well the proposed method generalizes to other contrasts, resolutions, and anatomies. Training with a data set that provides a wider range of imaging protocols could further improve performance. Training of such model is the subject of future work.

**Availability of data and material**

The authors thank the Intramural Research Program of the National Heart, Lung, and Blood Institute for the data obtained from the NIH Cardiac MRI Raw Data Repository. The reconstruction is shared in the Gadgetron repository  https://github.com/gadgetron/gadgetron.

**Funding**

**Authors' contributions**

--

**Table 1. Information for training and test datasets.**

| Category | Imaging application | Anatomy | Typical sequence parameters | Field strength | No. Samples and data format |
|---|---|---|---|---|---|
| Training and in-distribution testing | Retro-gated cine | Heart | Data acquisition with breath-holding<br>FOV: 360x270mm$^2$<br>Acquired matrix size: 256x144<br>Echo time: 1.28ms<br>Bandwidth: 977 Hz/pixel<br>Readout: SSFP<br>RF Flip angle: 50$^o$<br>Echo spacing: 2.97ms<br>Output phases: 30<br>Acceleration: R=2 | 3T | Training: N=7590 subjects, 96,605 cine series, 2,885,236 images, 61% male, mean age 54 years<br><br>Testing: N=231 subjects, 3,000 cine series, 89,899 images<br><br>2D+T time series<br>Input tensor: [B, 3, T, H, W] |
| Testing, out-of-distribution | Real-time cine | Heart | Data acquisition with single-shot free-breathing<br>FOV: 360x270mm$^2$<br>Acquired matrix size: 192x110<br>Echo time: 0.98ms<br>Echo spacing: 2.27ms<br>Bandwidth: 1100Hz/pixel<br>Readout: BSSFP<br>RF Flip angle: 50$^o$<br>Imaging duration: 39ms<br>Acceleration: R=5 | 1.5T | N=10 subjects, one slice per subject, 8 males, mean age 52 years<br><br>2D+T time series<br>Input tensor: [B, 3, T, H, W] |
| | Perfusion | Heart | Data acquisition with single-shot free-breathing<br>Contrast injection and dynamic contrast changes<br>Adenosine stress<br>FOV: 360x270mm$^2$<br>Acquired matrix size: 256x108<br>Echo time: 1.17ms<br>Single-shot TR: 80ms<br>Bandwidth: 850Hz/pixel<br>Readout: BSSFP<br>RF Flip angle: 50$^o$<br>Acceleration: R=4 | 1.5T | N=5 subjects, each had a stress and a rest scan, 3 slices per scan with 60 heart beats, 2 males, mean age 43 years<br><br>2D+T time series<br>Input tensor: [B, 3, T, H, W] |
| | Neuro | Brain | T1 MPRAGE sequence<br>FOV: 250x250mm$^2$<br>Acquired matrix size: 256x256<br>Echo time: 7.2ms<br>Bandwidth: 250Hz/pixel<br>Readout: Turbo spin echo<br>Echo spacing: 3.58ms<br>TI: 200ms<br>Acceleration: R=2x2 | 1.5T | N=1 male, 45 years old<br><br>3D imaging<br>Input tensor: [B, 3, D, H, W] |
| | Spine | Spine | T2 TSE sequence<br>FOV: 340x340mm$^2$<br>Acquired matrix size: 448x448<br>Echo time: 89ms<br>TR: 3000ms<br>Bandwidth: 260Hz/pixel<br>Readout: Turbo spin echo<br>Acceleration: R=2 | 1.5T | N=1 male, 45 years old<br><br>2D imaging for 15 slices<br>Input tensor: [B, 3, SLC, H, W] |

**Table 2. Results of in-distribution tests for two backbone types.**

| HRnet | #paras | SSIM | | | | PSNR | | | |
|---|---|---|---|---|---|---|---|---|---|
| | | Proposed | Without g-factor | Without MR noise | Without recon knowledge | Proposed | Without g-factor | Without MR noise | Without recon knowledge |
| CNNT-large | 54,678,306 | **0.70471** | 0.55817 | 0.37283 | 0.37647 | **54.89698** | 48.13811 | 41.31241 | 40.48224 |
| CNNT | 27,485,139 | 0.67622 | 0.57645 | 0.37639 | 0.37533 | 54.13667 | 49.19455 | 41.51258 | 40.38141 |
| Swin3D | 54,664,836 | 0.67898 | 0.58599 | 0.402 | 0.39727 | 53.77681 | 49.54128 | 42.53329 | 41.23401 |
| ViT3D | 27,478,404 | 0.62632 | 0.59992 | 0.54525 | 0.47964 | 51.75421 | 49.88557 | 47.9071 | 44.0014 |
| Conv3D | 22,815,891 | 0.60142 | 0.57271 | 0.56566 | 0.46434 | 50.74263 | 49.05661 | 48.72905 | 41.51646 |
| ViT2D | 17,746,308 | 0.47747 | - | - | - | 46.95545 | - | - | - |
| Conv2D | 20,382,867 | 0.46782 | - | - | - | 46.67031 | - | - | - |

| Unet | #paras | SSIM | | | | PSNR | | | |
|---|---|---|---|---|---|---|---|---|---|
| | | Proposed | Without g-factor | Without MR noise | Without recon knowledge | Proposed | Without g-factor | Without MR noise | Without recon knowledge |
| CNNT-large | 48,880,418 | 0.69784 | 0.54617 | 0.37882 | 0.37367 | 54.70398 | 47.65097 | 41.52626 | 40.31814 |
| CNNT | 25,226,195 | 0.67355 | 0.57025 | 0.37785 | 0.37749 | 54.09462 | 48.9332 | 41.51152 | 40.48785 |
| Swin3D | 49,309,316 | 0.62817 | 0.48382 | 0.50779 | 0.3971 | 51.58844 | 45.2311 | 46.46576 | 41.21689 |
| ViT3D | 25,661,828 | 0.62002 | 0.60346 | 0.57839 | 0.48495 | 51.25344 | 50.48932 | 49.30984 | 44.25077 |
| Conv3D | 18,787,475 | 0.62299 | 0.58958 | 0.49493 | 0.43549 | 51.78548 | 50.02364 | 46.15332 | 40.96275 |
| ViT2D | 15,487,364 | 0.46882 | - | - | - | 46.53302 | - | - | - |
| Conv2D | 16,206,995 | 0.49678 | - | - | - | 47.43377 | - | - | - |

**Table 3. Results for real-time cine and perfusion generalization tests.**

| Tests | Measurements | ROIs | *Raw | *Proposed | ++Without g-factor | §Without MR noise | ‡Without recon knowledge | P values | |
|---|---|---|---|---|---|---|---|---|---|
| **Real-time Cine, R=5** | SNR | Blood pool<br><br>G-factor<br>4.053±<br>1.039 | 13.474±<br>4.846 | 70.043±<br>11.704 | 22.194±<br>7.880 | 19.326±<br>6.781 | 18.430±<br>5.979 | + vs. * | 1.6936e-08 |
| | | | | | | | | + vs. ++ | 2.5726e-08 |
| | | | | | | | | + vs. § | 3.0672e-08 |
| | | | | | | | | + vs. ‡ | 3.0949e-08 |
| | | Myocardium<br><br>G-factor<br>3.947±<br>0.795 | 5.806±<br>2.200 | 20.395±<br>3.220 | 7.394±<br>2.700 | 8.088±<br>2.909 | 7.639±<br>2.725 | + vs. * | 2.166e-07 |
| | | | | | | | | + vs. ++ | 7.616e-07 |
| | | | | | | | | + vs. § | 1.7574e-06 |
| | | | | | | | | + vs. ‡ | 1.1284e-06 |
| | CNR | Blood pool and Myocardium | 7.668±<br>2.981 | 49.648±<br>9.804 | 14.800±<br>5.493 | 11.238±<br>4.181 | 10.791±<br>3.807 | + vs. * | 9.2284e-08 |
| | | | | | | | | + vs. ++ | 1.1031e-07 |
| | | | | | | | | + vs. § | 1.2977e-07 |
| | | | | | | | | + vs. ‡ | 1.8059e-07 |
| **Perfusion, R=4** | SNR | Blood pool<br><br>G-factor<br>1.913±<br>0.516 | 24.542±<br>14.805 | 74.054±<br>26.319 | 70.097±<br>31.543 | 59.048±<br>32.052 | 46.535±<br>23.578 | + vs. * | 9.2301e-17 |
| | | | | | | | | + vs. ++ | 0.011635 |
| | | | | | | | | + vs. § | 1.5111e-06 |
| | | | | | | | | + vs. ‡ | 4.2334e-15 |
| | | Myocardium<br><br>G-factor<br>1.870±<br>0.436 | 4.514±<br>2.688 | 16.689±<br>5.861 | 12.497±<br>6.246 | 9.959±<br>5.528 | 8.979±<br>5.208 | + vs. * | 5.5756e-16 |
| | | | | | | | | + vs. ++ | 8.2579e-09 |
| | | | | | | | | + vs. § | 8.5671e-12 |
| | | | | | | | | + vs. ‡ | 7.8484e-12 |
| | CNR | Blood pool and Myocardium | 20.028±<br>12.419 | 57.365±<br>24.628 | 56.601±<br>27.275 | 49.089±<br>27.503 | 37.557±<br>19.432 | + vs. * | 2.6235e-13 |
| | | | | | | | | + vs. ++ | 0.84282 |
| | | | | | | | | + vs. § | 0.0014682 |
| | | | | | | | | + vs. ‡ | 4.0298e-11 |

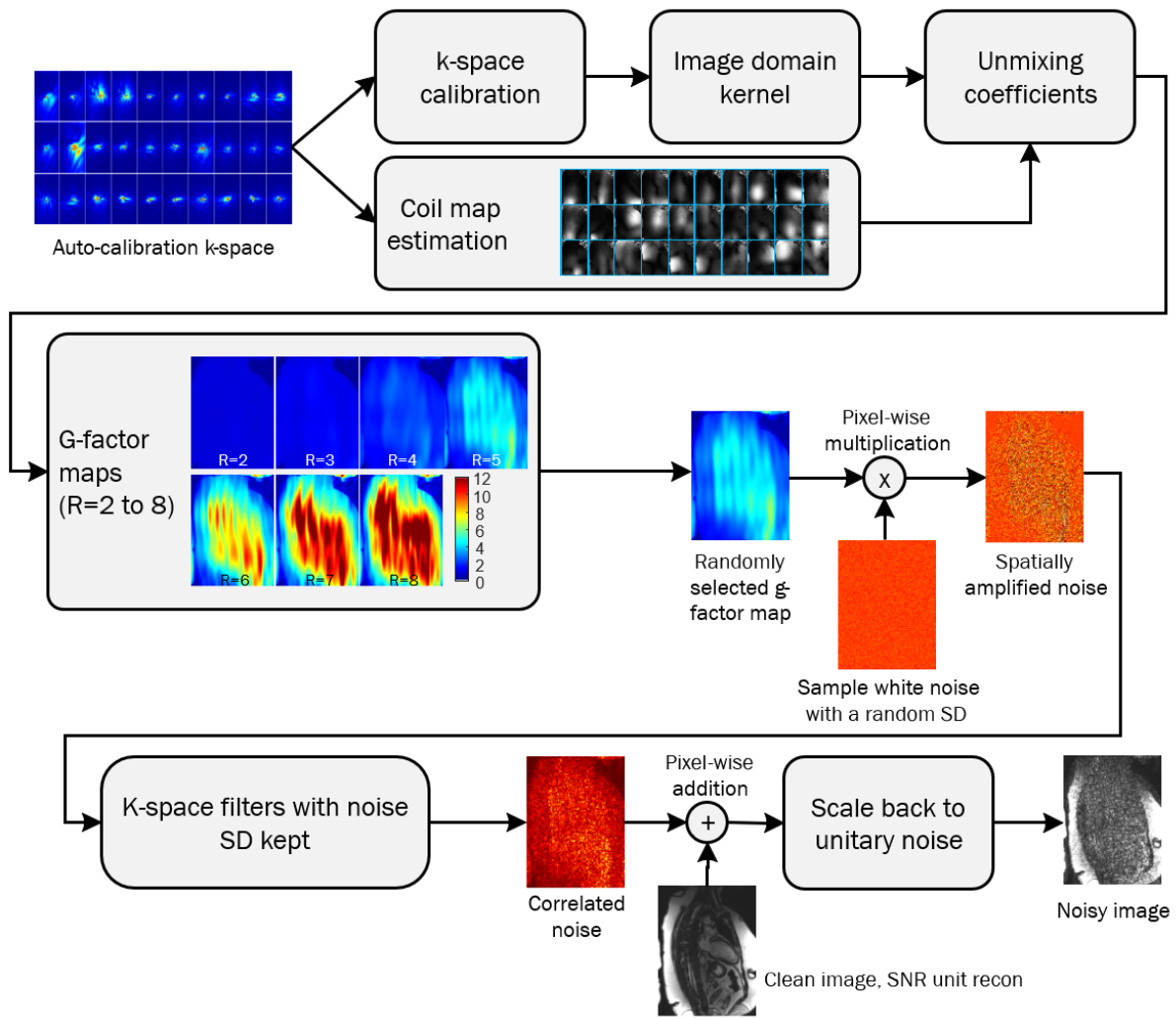

(a) Create training data and G-factor map augmentation

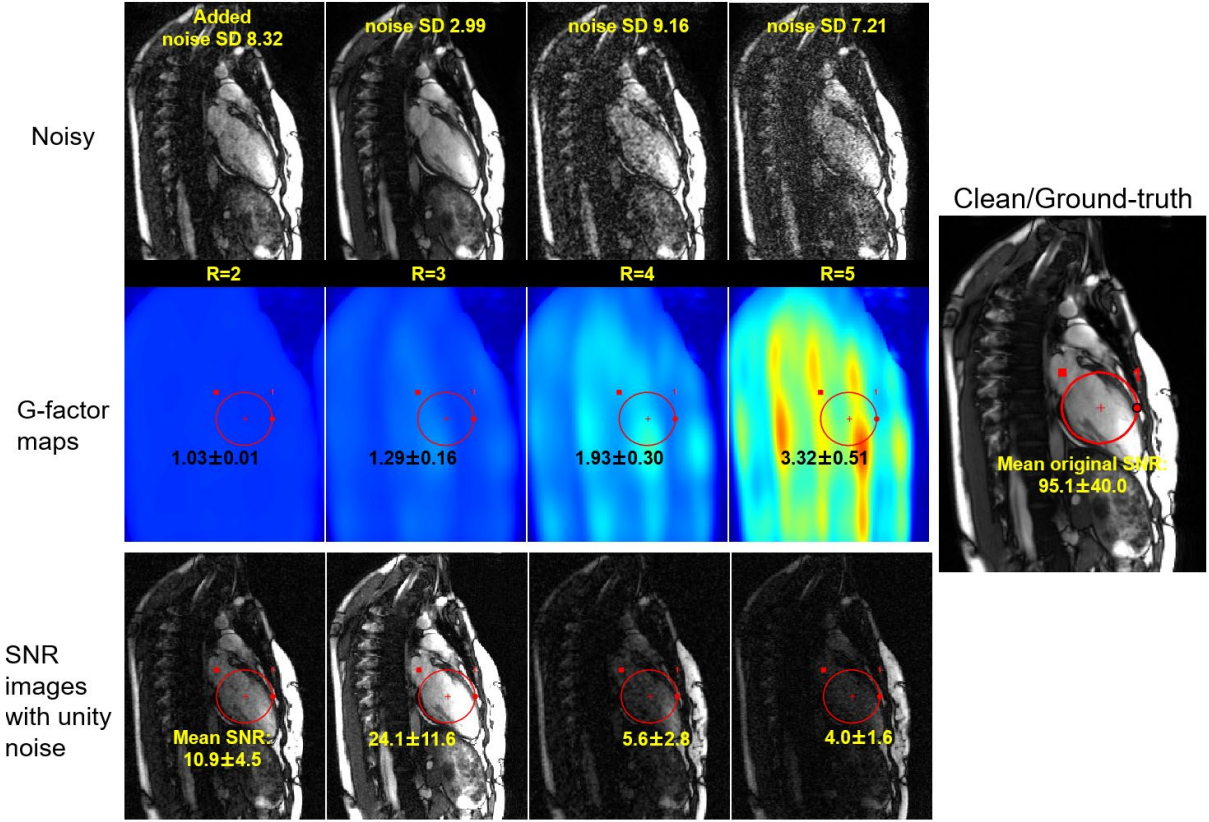

(b) Examples of training samples with a wide range of SNR

**Figure 1**. Training sample creation. (a) The training samples are paired clean and noisy image series. The noisy images are created by computing the g-factor maps and using them to generate noise images. Geometry (g)-factor maps were computed from the auto-calibration *k*-space. GRAPPA calibration was computed for accelerations R=2 to 8. The GRAPPA k-space kernel was converted to image domain kernel and unmixing coefficients were computed by combining image domain kernels and coil maps. For data augmentation, a g-factor map is randomly selected and pixel-wise multiplied to the white noise. The resulting spatially varying noise further goes through k-space filtering steps to introduce correlation. The final noise is added to the clean image and scaled to be unitary, as the noisy sample for training. (b) Four noisy samples are created from a clean cine series. The original SNR is 95.1 in the ROI. By randomly selecting a g-factor map and changing the starting noise level, a wide range of SNRs can be produced. The SNR images are computed by dividing the images by the g-factor map.

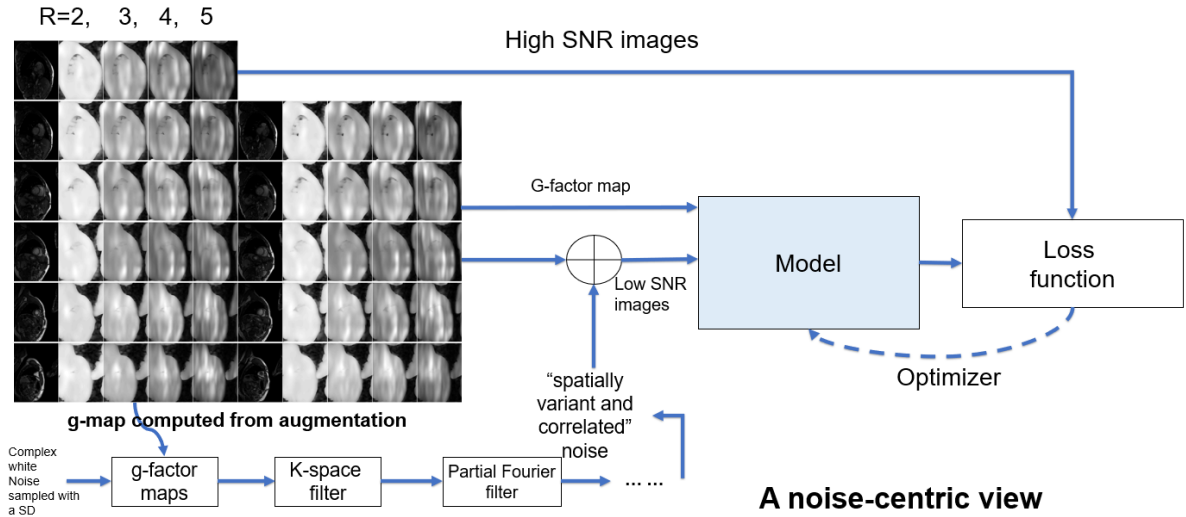

(a) Training scheme

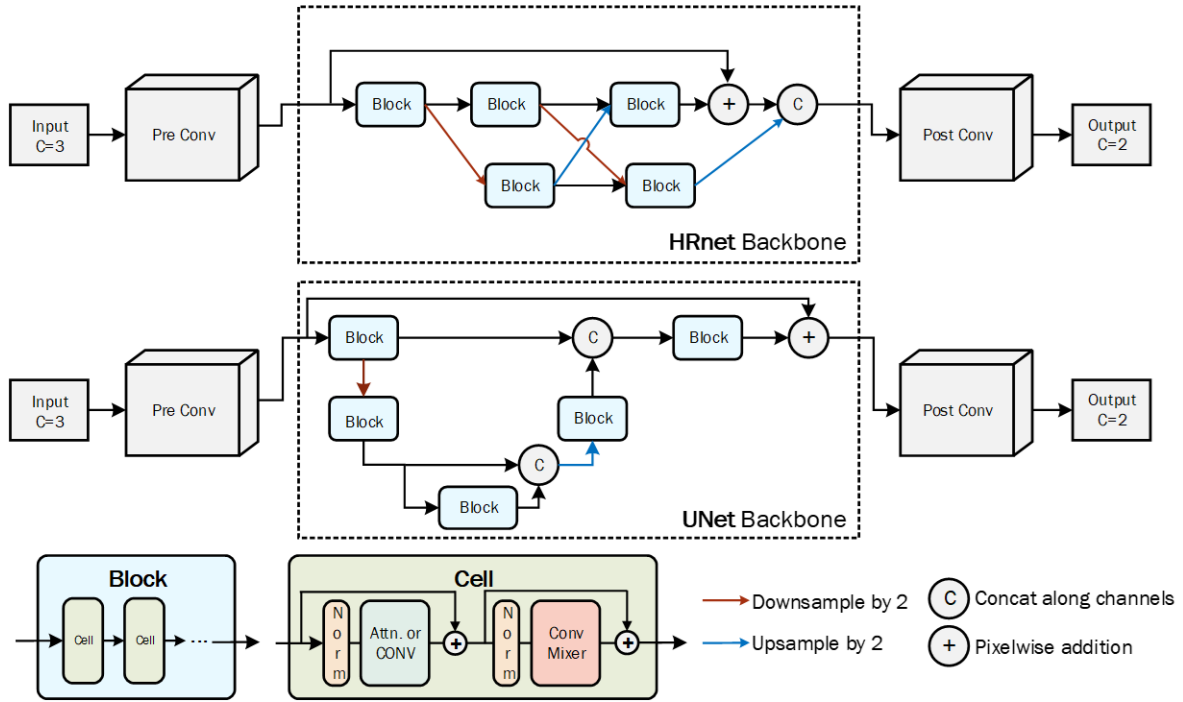

(b) Model design

**Figure 2**. Overview of training scheme and model design. (a) Reconstructed cine images are augmented with spatially varying and correlated noise to create noisy samples. The corresponding g-factor maps are concatenated to the images and used as input into the model. The model predicts high SNR images. (b) All models consist of a pre-conv layer as the shallow feature extractor, a backbone and the output convolution. To simplify the evaluation on different models, a cell-block-backbone design is proposed for the backbone. Two backbone architectures tested here are HRnet and Unet. Both backbones process tensors through blocks which are connected by the downsample/upsample operation. Every block consists of 3 to 6 cells. Every cell follows the standard design, including normalization, attention or convolution layers and skip connections. By changing the module in the cell, different transformer and convolution models are instantiated and tested.

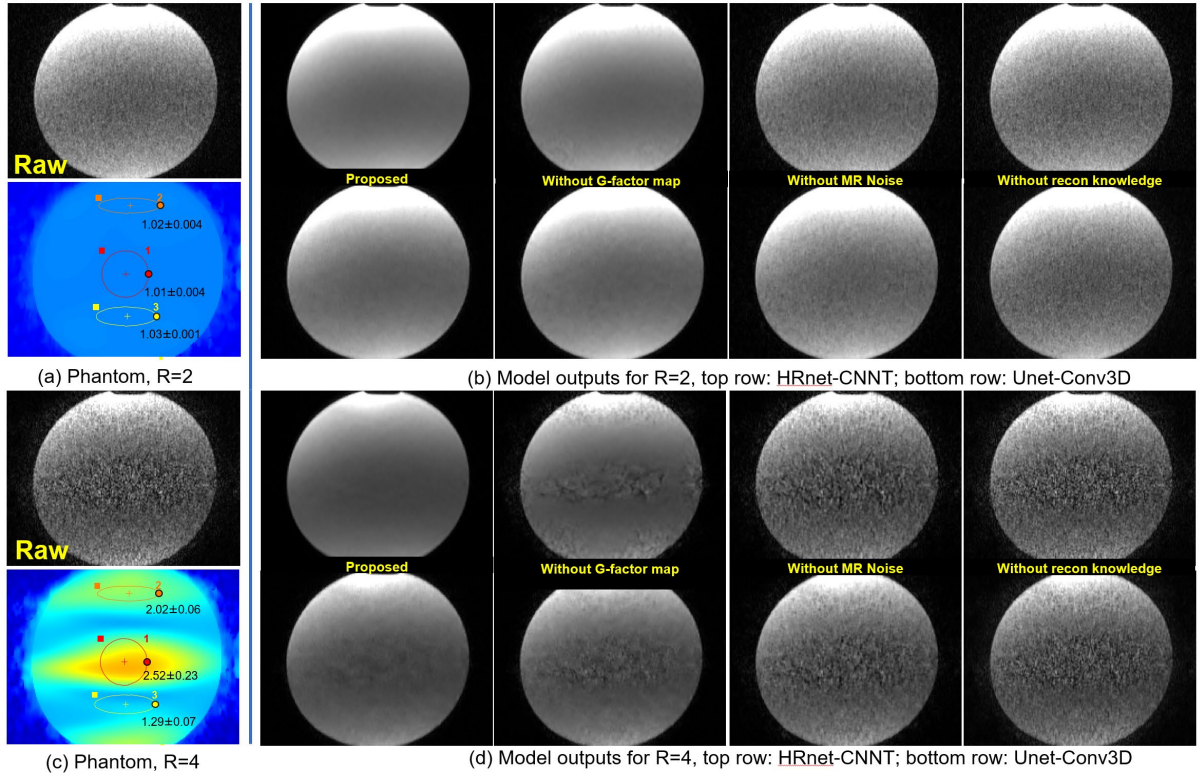

**Figure 3.** Phantom test results. (a) The raw images of R=2 acceleration and g-factor map show a very minor noise amplification. (b) Model outputs with one transformer and one convolution architecture for proposed training, compared to ablation tests. (c) The images and g-factor map for R=4 acquisition shows lower SNR and spatial noise amplification. The g-factor is higher at 2.52 at the center of FOV. The proposed method removes noise amplification. Training without g-factor map results in less efficient noise removal. For both accelerations, transformer models outperform convolution. Training without MR noise distribution further degrades the performance.

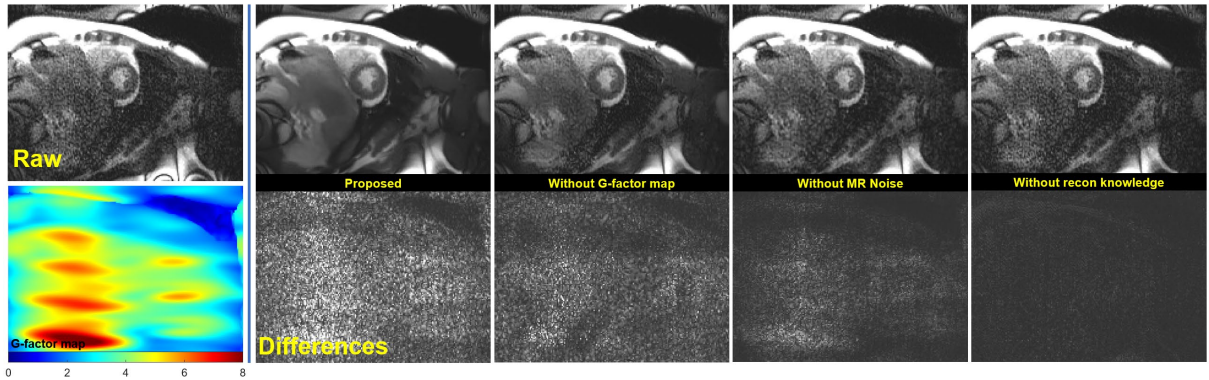

**Figure 4**. Real-time cine (acceleration R=5) results produced with the HRnet-CNNT-large model. The raw SNR is lower with elevated spatial noise amplification due to acceleration. The proposed training method produced the best results.

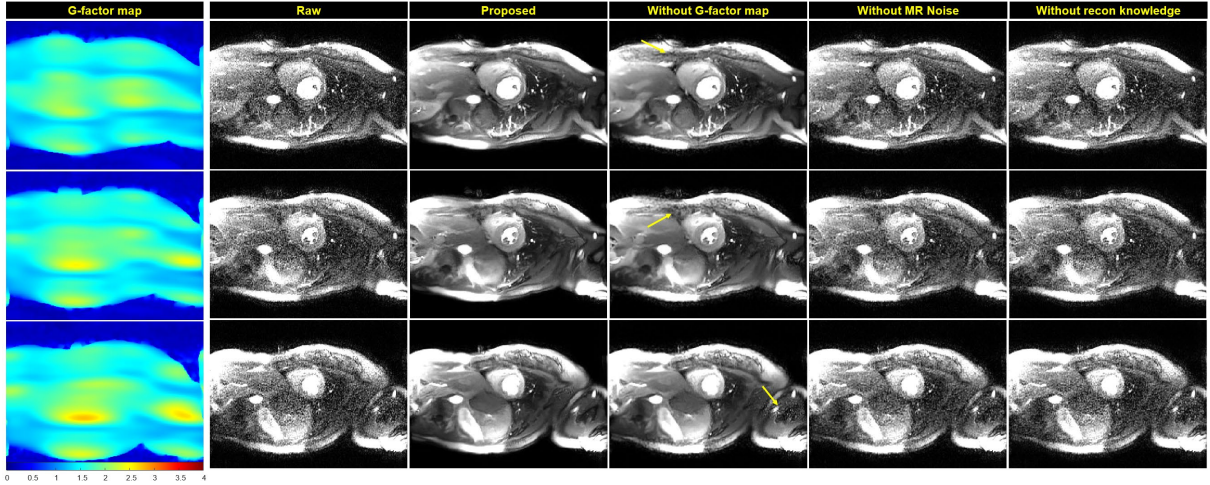

**Figure 5**. Result for accelerated (R=4) myocardial perfusion imaging. The contrast passage creates dynamically varying contrast which was not seen in the training dataset. Moreover, the saturation preparation reduced the base SNR. Despite the out-of-distribution challenges, the model generalized well to perfusion imaging. Similarly to previous tests, noise reduction is more effective when knowledge about noise distribution is included in training.

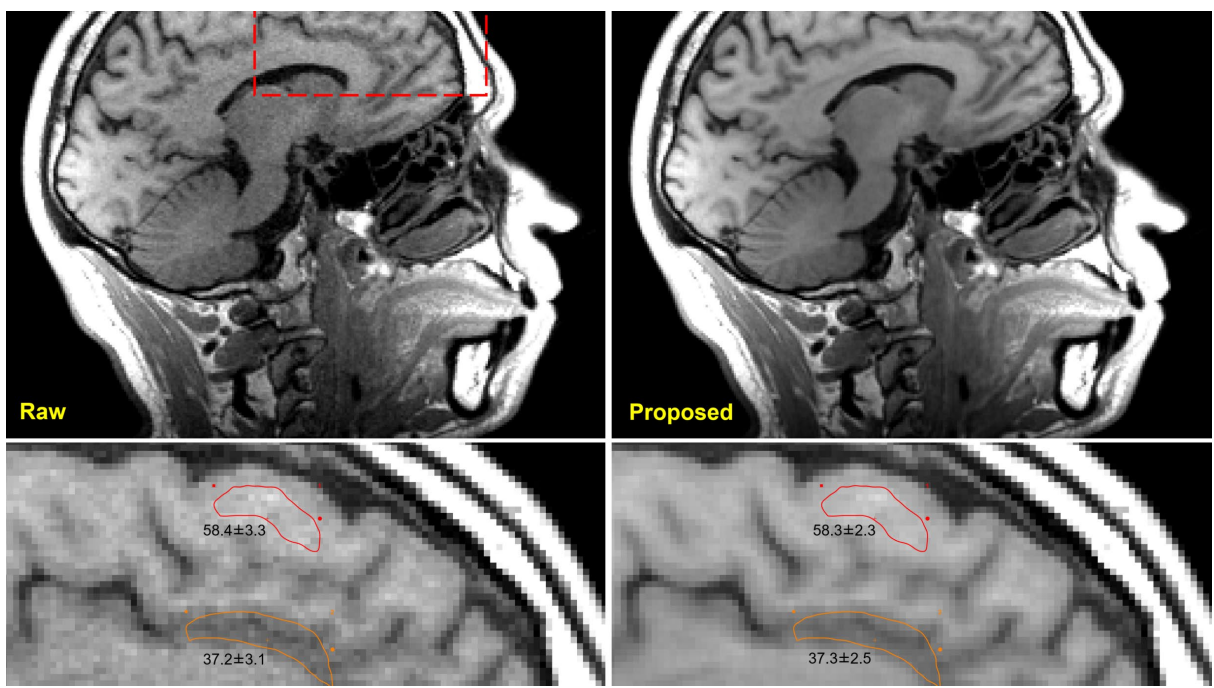

**(a) T1 MPRAGE neuro scan**

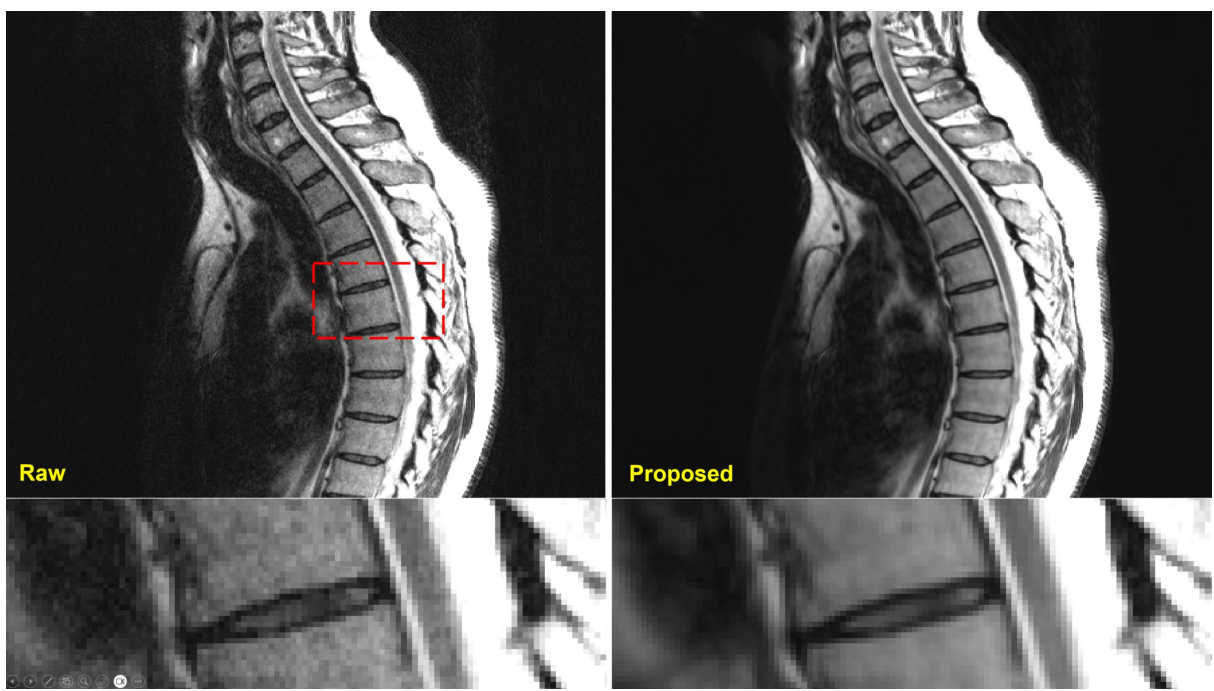

**(b) T2 TSE spine scan**

**Figure 6.** Generalization tests for different anatomies. (a) A R=2x2 MPRAGE T1 neuro scan was acquired, reconstructed and processed with trained model. The training dataset did not include any neuro data, yet the model generalized well, with noticeable SNR improvement and preserved gray-white matter contrast. (b) A R=2 T2 TSE spine scan was processed with the trained model. Training data did not include spine scans and did not include the high spatial resolution ($0.76mm^2$) of this acquisition. The model generalized well to this application regardless.

**Supplemental Appendices**

**Appendix E1. Information for deep learning models**

As shown in Figure 2, the model consists of three components: pre-convolution layer, backbone and post-convolution layer. The input tensors are in the shape of [B, C, T/S/D/Z, H, W]. C is 3 for complex inputs (real, imagery and g-factor). Noise in the input images are scaled to 1.0×g-factor, as this setup is consistent with reconstruction outputs.

The pre-convolution layer is a shallow feature extractor (36). It is kept being minimal as a 2D convolution to uplift input channel C to 64, encouraging backbone to take on most heavy lifting and helping generalization. The post-convolution is another CONV layer, converting $C_{backbone}$ after the backbone to required output channels (2 for complex training and 1 for magnitude training).

Two well-known backbone architectures, HRnet and Unet, are implemented and tested in this study. Both architectures utilize the multi-resolution pyramid to balance model size, expressive power and computing cost. The building components include multiple Blocks, downsample and upsample layers, channel-wise concatenation, and skip connection. The HRnet maintains a longer pipeline on the original tensor size and Unet is smaller in size and less computing expensive.

The input tensors are processed through every block, gaining more channels and reducing spatial resolution, which is explained by the backbone plots annotated with tensor sizes.

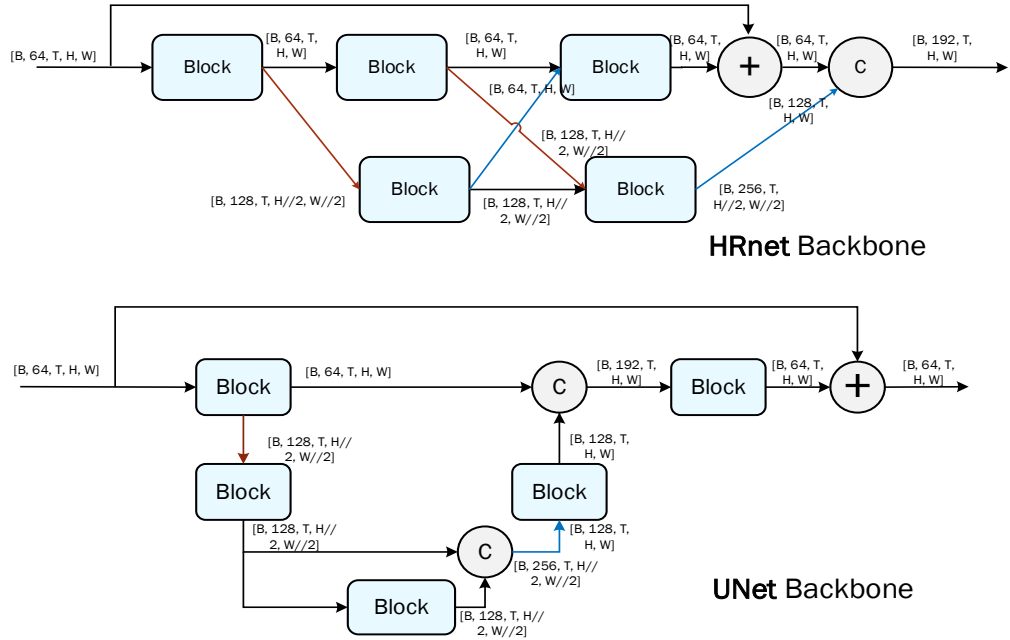

**Figure 1.** Annotated backbone architectures.

Downsampling was implemented with patch merging (24) followed by a convolution to format outputs to have the required number of channels. The upsampling was implemented with a linear interpolation followed by a CONV layer.

Backbones consist of several blocks. A block is a container of N cells. Every cell has a classical setup of two skip connections, layer norms (37) and attention or convolution layers.

By switching the attention methods (e.g. Swin3D, ViT3D or CNNT etc.), we can instantiate different models for experiments. A pure convolution model was implemented by replacing attention with convolution layers.

Every block in all models, except CNNT-large, has 3 cells. For CNNT-large, a block holds 6 blocks. By inserting more cells or more blocks, the model can be scaled up or down.

As used in other denoising training schemes, models were trained on image patches to encourage models to focus on noise distribution instead of image content. The patch size was [T/S/D/Z=16, H=64, W=64]. The window size in Swin3D and ViT3D was [16, 8 ,8], where every [2, 2, 2] neighborhood was processed as a token. The CNNT transformer method computed attention between all [H, W] frames without explicit neighborhood tokenization. All convolutions had the kernel size 3 and padding 1. We note that unlike the original Swin and ViT papers, we re-patch and un-patch the tensors before and after every operation, resulting in imaging tensors that can be processed by the normalization and convolutional mixer layers in every cell.

# Supplemental Data

**Movie 1**: The movies correspond to the example in Figure 1b. The ground-truth clean image is the single one on the left. The first row are the noisy samples. The second row are the SNR images.

**Movie 2**: Corresponding movies to Figure 4 are given here.

**Movie 3**: More R=5 real-time cine examples are given here. In all cases, proposed training noticeably improves performance. The leftover noise amplification is very visible without the g-factor map.

**Movie 4**: Movies of perfusion denoising corresponding to Figure 5 are presented. Model generalized well to dynamic contrast and low base SNR.

**Movie 5**: Movie corresponds to Figure 6a for the T1 MPRAGE neuro test.

**Movie 6**: Movie corresponds to Figure 6b for the T2 TSE spine test.